\documentclass[a4paper,12pt]{article}
\usepackage{epsfig} 
\usepackage{amsmath}

\newcommand{\beq}{\begin{equation}}
\newcommand{\eeq}{\end{equation}}
\newcommand{\beqa}{\begin{eqnarray}}
\newcommand{\eeqa}{\end{eqnarray}}
\newcommand{\nn}{\nonumber}

\newcommand{\half}{\frac{1}{2}}

\newcommand{\we}{\wedge}

\newcommand{\der}{\partial}

\begin{document}

\author{%
Monika Pietrzyk and Igor Kanatchikov \\
School of Physics and Astronomy, University of St Andrews\\
North Haugh, KY16 9SS, St Andrews, UK\\
}
\title{Polysymplectic Integrator \\
for the Short Pulse Equation} 

\maketitle

\begin{abstract}
The polysymplectic analysis of the Short Pulse Equation known in nonlinear 
optics is used in order to construct a geometric polysymplectic integrator 
for it. The proposed scheme turn out to be much more effective than other 
standard integration schemes for nonlinear PDEs, such as the pseudo-spectral
integrator. In our numerical experiments the polysymplectic integrator 
appears to be an order of magnitude more precise and approximately 
2.5 times faster at long propagation times 
than  the pseudo-spectral method. 
\end{abstract}


\section{Introduction}\label{aba:sec1}

The multisymplectic Hamiltonian formalism has emerged from 
geometric theories in the calculus of variations\cite{gimmsy}. 
It has been a subject of numerous investigations recently
\cite{sardan,deleon1,deleon2,paufler,francav,romanroy,krupka,romp98}. 
The polysymplectic formulation was proposed as a certain version 
of it, which allows to define proper Poisson brackets 
for the purpose of  field quantization
\cite{bial97,ijtp2001,opava2001,ym}. The multisymplectic approach 
to the construction of geometric numerical integrators of PDEs  
was proposed in \cite{marsden98}. The application of the closely related 
``multi-symplectic'' structure in wave propagation has been pioneered by 
Bridges\cite{bridges97}. 

In this contribution we apply the polysymplectic formalism to the short 
pulse equation (SPE) known in nonlinear optics. The short pulse equation 
has been proposed \cite{sp1,sp2} as a description of few-cycle pulses 
when the standard nonlinear Schr\"odinger equation cannot be applied because 
the slowly varying envelope approximation it is based on becomes questionable. 
In \cite{sakovich1,brunelli} the integrability of this equation has been 
proven, and in \cite{sakovich2} an example of the exact solution has been 
constructed. In \cite{wias06} three integrable two component generalizations 
of SPE have been found.  

Here we apply  the polysymplectic formalism in order to construct a 
polysymplectic geometric integrator for SPE. This work is a part of the 
investigation of the properties of ultra-short pulses in nonlinear optics 
with the help of SPE and its generalizations which requires a stable 
and robust numerical integration scheme for SPE.  

The polysymplectic formulation of SPE is discussed in Sect. 2. In Sect. 3 
we construct the simplest polysymplectic integrator and briefly compare its 
effectiveness  with the well known pseudo-spectral numerical integration 
\cite{Fornberg}. 

\section{The polysymplectic formulation of SPE} 

The short pulse  equation 
\beq
u_{xt} = u + \frac{1}{6}(u^3)_{xx} 
\eeq
can be written in the form 
\beq
\phi_{xt} - \phi - \frac{1}{6} (\phi_x^3)_x =0 
\eeq
if we introduce the potential $\phi$
\beq
u := \phi_x .
\eeq
This equation can be derived  from the first order Lagrangian 
\beq
L=\half \phi_t\phi_x - \frac{1}{24} \phi_x^4 +\half \phi^2 .  
\eeq
Using the standard polysymplectic 
(De Donder-Weyl) Hamiltonian formalism, we introduce the polymomenta  
\beqa 
&&p^t := \frac{\der L}{\der \phi_t } = \half\phi_x ,\nn\\
&&p^x := \frac{\der L}{\der \phi_x } =  \half\phi_t - \frac{1}{6} \phi_x^3 , 
\eeqa
and the (De Donder-Weyl) Hamiltonian 
\beq
H_{DW}:= p^t\phi_t + p^x\phi_x - L 
= 2 p^x p^t +\frac{2}{3} (p^t)^4 - \half\phi^2 . 
\eeq
Then the polysymplectic (De Donder-Weyl) Hamiltonian equations take the form 
\beqa 
&&\der_x p^x + \der_t p^t = - \frac{\der H}{\der \phi}  = \phi , \nn \\
&& \der_x \phi =  \frac{\der H}{\der p^x} = 2p^t , 
\label{Donder-Weyl} \\
&& \der_t \phi =  \frac{\der H}{\der p^t} = 2p^x + \frac{8}{3} (p^t)^3 . \nn 
\eeqa
This set of first order equations is equivalent to SPE written 
in terms of the potential function $\phi(x,t)$, Eq. 2.  
It is well known that these equations can be obtained from the geometrical 
formulation of first order variational problems using the Poincare-Cartan form 
and its exterior derivative 
 \cite{gimmsy,romp98}.
\beq
\Omega = d \phi \we dp^x \we dt + d\phi \we dp^t \we dx - dH\we dx \we dt . 
\eeq

In order to establish a  connection with the multi-symplectic formulation of 
Bridges\cite{bridges97}  which has became more popular in discussions of 
geometric integrators of PDEs, let us introduce the set of variables 
$Z^v := (\phi, p^x, p^t)$. 
Then the DW Hamiltonian equations can be written in matrix form 
\beq
\beta^x\der_x Z + \beta^t\der_t Z = \nabla_Z H ,
\eeq
where the $\beta$-matrices 
\beq
\beta^x = 
\left( 
\begin{array}{ccc}
0 & 0 & -1 \\
0 & 0 & 0 \\
1 & 0 & 0 \end{array} 
\right) , \quad 
\beta^t = 
\left( 
\begin{array}{ccc}
0 & 0 & -1 \\
0 & 0 & 0 \\
1 & -0 & 0 \end{array} 
\right) , 
\eeq 
can be identified with the so-called Duffin-Kemmer-Petiau matrices (in 2D) 
\cite{dkp99}
which fulfill the DKP algebra relations $(a,b,c = (x,t))$.
\beq
\beta^a \beta^b \beta^c + \beta^c \beta^b \beta^a = - \beta^a \delta^{bc} 
- \beta^c \delta^{ab} .  
\eeq
This form of DW Hamiltonian equations generalizes the Hamiltonian equations 
in mechanics written in the form 
\[ \omega \der_t Z = \der_Z H ,  \] 
where 
 $\left( 
\begin{array}{cc}
 0 & 1 \\
 -1 & 0  \end{array} 
\right) $ is the symplectic matrix and $Z:= (p,q)$ . 
 
Associated with the above two anti-symmetric matrices $\beta $ 
are two components of the polysymplectic form 
\beqa 
&&\kappa^x = \half dz \we \beta^x dz = -  dp^x\we d\phi , \nn\\
&&\kappa^t = \half dz \we \beta^t dz = dp^t\we d\phi .
\eeqa
The structure given by two components of the polysymplectic form $\kappa^x$ and $\kappa^t$ 
was called multi-symplectic by Bridges \cite{bridges97}. In the notations 
introduced by Bridges (1997) $\beta^x = K$ and $\beta^t = M$ and $H = -S$.   
These  notations are now standard in the papers devoted to the geometric 
(multisymplectic) integrators of PDEs 
\cite{bridges-reich,zhao,chen,frank,future}. 
In this notation the fundamental {\it multi-symplectic conservation law } is 
written in the form: 
 \beq
d/dt \kappa^t + d/dx \kappa^x = 0 .
\eeq 
It is equivalent to the on-shell exactness of the polysymplectic form. 

\section{Polysymplectic integrator for SPE } 

The simplest realization of the polysymplectic integrator  
is constructed by  the discretization of DW Hamiltonian equations using 
the midpoint method in both $x$ and $t$ directions. Using the following 
definitions:
\beqa
&&\phi_{i,j} \approx \phi(i\Delta x, j \Delta t ), \quad 
\nn \\
&&\phi_{i+1,j+1/2} := \half (\phi_{i,j} + \phi_{i,j+1}  ) , 
\\
&& \phi_{i+1/2,j+1/2} := \frac{1}{4} (\phi_{i,j} + \phi_{i,j+1}+ \phi_{i+1,j} 
+ \phi_{i+1,j+1}   ) . \nn
\eeqa
and the derivatives are expressed by:
\beqa
\delta_x \phi_{i,j} := \frac{\phi_{i+1,j+1/2}-\phi_{i,j+1/2}}{\Delta x},
\quad
\delta_t \phi_{i,j} := \frac{\phi_{i+1/2,j+1}-\phi_{i+1/2,j}}{\Delta x} 
\eeqa

\noindent 
we can write the discretized version of the polysymplectic formulation 
of the SPE (eq. \ref{Donder-Weyl}):
\begin{subequations}
\beqa
 &&\frac{p^x_{i+1,j+\half}- p^x_{i,j+\half}}{\Delta x} 
+  \frac{p^t_{i+\half, j+1} -p^t_{i+\half, j}}{\Delta t} 
= \phi_{i+\half,j+\half} , \label{discrete_a}
\\
&&\frac{\phi_{i+1,j+\half}- \phi_{i,j+\half}}{\Delta x} 
= 2p^t_{i+\half,j+\half}  , \\
&& \frac{\phi_{i+\half, j+1} -\phi_{i+\half, j}}{\Delta t} 
= 2p^x_{i+\half,j+\half} + \frac{8}{3} (p^t_{i+\half,j+\half})^3 .
\label{discrete_c}
\eeqa 
\end{subequations}\

Making simple calculations one can prove (see also \cite{bridges-reich})
that the discrete version of the 
polysymplectic formulation of the SPE fully satisfies the discrete version 
of the polysymplectic conservation law:
\beqa
\frac{\kappa^t_{i+1/2,j+1}-\kappa^t_{i+1/2,j}}{\Delta t}+ 
\frac{\kappa^x_{i+1,j+1/2}-\kappa^x_{i,j+1/2}}{\Delta x},
\eeqa
where
$\kappa^t_{i+1/2,j} = d p^t_{i+1/2,j} \wedge \phi_{i+1/2,j}$
and 
$\kappa^x_{i,j+1/2} = d p^t_{i,j+1/2} \wedge \phi_{i,j+1/2}$.

\subsection{The numerical implementation}

We will test our numerical polysymplectic integrator using the exact soliton 
solutions of the SPE. We solve the initial boundary value problem, 
$u(x,t=0)=u0$, which discretized gives $u_{i,j=0} = u0_i, i=1,...,N$. 
We also assume that the values of the solution vanishes on the right 
boundary, $u_{N,j} = 0, j=0,1,2...$ (the wave propagates from the right 
to the left). By straightforward calculations we can obtain the initial
and boundary values of polysymplectic variables $p^t_{i,j=0}$, $\phi_{i,j=0}, 
i=1,N$, and $p^t_{N,j}= \phi_{N,j} =p^x_{N,j}+ p^x_{N,j+1} = 0, j=0,...,M $. 

Knowing values of polysymplectic variables $p^t, p^x$ and $\phi$ at the 
three mesh points $(i+1,j)$, $(i+1,j+1)$, and $(i,j)$ (see Fig. 1) we 
calculate values of polysymplectic variables at the grid point $(i,j+1)$, 
namely given $p^t_{i+1,j}$, $p^t_{i+1,j+1}$, $p^t_{i,j}$, $\phi_{i+1,j}$,
$\phi_{i+1,j+1}$, $\phi_{i,j}$, and  $p^x_{i+1,j+1} + p^x_{i+1,j}$, we 
calculate $p^t_{i,j+1}$, $\phi_{i,j+1}$ and $p^x_{i,j+1} + p^t_{i,j}$. 

This can be done by manipulating the set of three nonlinearity coupled 
polysymplectic discrete equations (\ref{discrete_a}-\ref{discrete_c}),
which gives: \begin{subequations}
\[
(p^t_{i,j+1})^3 
+ 3 (p^t_{i+1,j}+p^t_{i,j}+p^t_{i+1,j+1}) (p^t_{i,j+1})^2
\]
\[
+ 3 \left( (p^t_{i+1,j}+p^t_{i,j}+p^t_{i+1,j+1})^2 + 
4 \left(\frac{2 \Delta x}{\Delta t} + \frac{(\Delta x)^2}{2}\right) \right)
p^t_{i,j+1}
\]
\[
-\frac{24}{\Delta t}(\phi_{i+1,j+1}-\phi_{i,j})
-12 \Delta x( \phi_{i+1,j}+\phi_{i+1,j+1})
\]
\[
+12 \left(\frac{2 \Delta x}{\Delta t} + \frac{(\Delta x)^2}{2}\right) 
p^t_{i+1,j+1} 
+ 6 (\Delta x)^2 (p^t_{i+1,j}+p^t_{i,j})
\]
\begin{equation}
+24 (p^x_{i+1,j} + p^x_{i+1,j+1})
+ (p^t_{i+1,j} + p^t_{i,j} + p^t_{i+1,j+1})^3 =0 , 
\label{diska}
\end{equation}
\vspace*{3mm}
\begin{equation}
\phi_{i,j+1} = (\phi_{i+1,j} - \phi_{i,j} + \phi_{i+1,j+1})
- \Delta x \, (p^t_{i+1,j} + p^t_{i,j} + p^t_{i+1,j+1})
- \Delta x \, p^t_{i,j+1} , 
\label{diskb}
\end{equation}
\vspace*{3mm}
\[
(p^x_{i,j+1}+p^x_{i,j}) = (p^x_{i+1,j+1}+p^x_{i+1,j}) 
- \frac{\Delta x}{\Delta t} (p^t_{i,j} + p^t_{i+1,j} - p^t_{i+1,j+1})
\]
\begin{equation}
+ \frac{\Delta x}{\Delta t} p^t_{i,j+1}
- \frac{\Delta x}{2} (\phi_{i+1,j} + \phi_{i,j} + \phi_{i+1,j+1}
- \frac{\Delta x}{2} \phi_{i,j+1} .
\label{diskc}
\end{equation}
\end{subequations}

\begin{figure}
\hspace*{5mm}\psfig{file=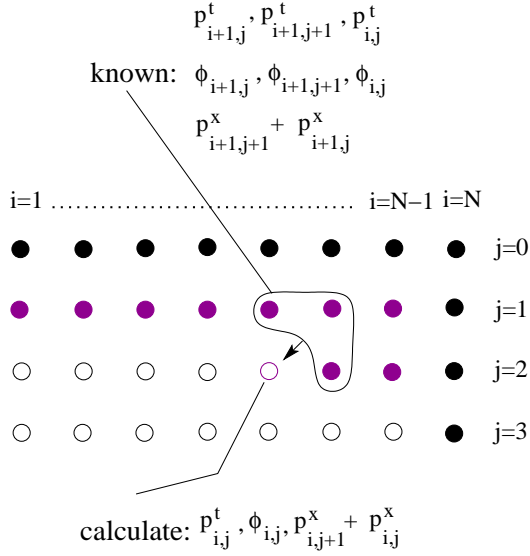,width=0.5\linewidth}
\caption{The discretization mesh.}
\end{figure}

\vspace{5mm}
\begin{figure}
\psfig{file=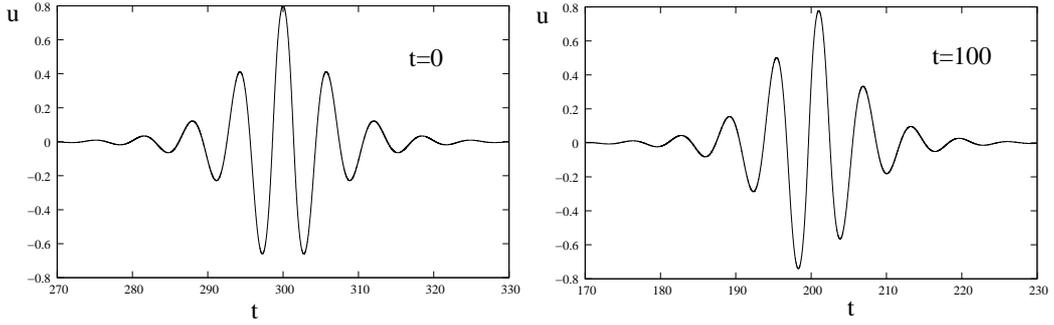,width=1.\linewidth}
\caption{The evolution of the Sakovich' solution (with $m=0.2$)
for $t=0$ and $t=100$. }
\end{figure}

\vspace*{3mm}
Using the cubic Eq. (\ref{diska}) we first calculate $p^t_{i,j+1}$ (we
select only the root which ensures the continuity of the solution). 
Then  Eqs. (\ref{diskb}) and (\ref{diskc}) yield, respectively, 
$\phi_{i,j+1}$ and $p^x_{i,j+1} + p^x_{i,j}$ (see Fig. 1). 

We can now transfer back the polysymplectic variables to the amplitude of 
the electric field and knowing $u_{i+1,j} \, u_{i,j}, \, u_{i+1,j+1}$
we can calculate $u_{i,j+1} = 2 p^t_{i,j+1}$.

\vspace{5mm}

In order to test the effectiveness of the method, we 
numerically propagate the known  Sakovich' exact solution of SPE 
\cite{sakovich2}
to $t=100$. The evolution of the Sakovich exact solution (with $m=0.2$) 
is shown on Fig. 2 at $t=0$ and $t=100$.

The exact solution is compared with the numerical solutions obtained using 
the polysymplectic scheme and the pseudo-spectral scheme. We compare the 
error of the methods and the CPU time required to reach $t=100$ at different 
values of discretization steps $\Delta t$ and $\Delta x$. The error of 
numerical integration is given by the standard deviation:
\begin{equation}
\sigma= 
\sqrt{\frac{1}{N} \sum_{i=1}{N} \left(u_{i,j} - \bar{u}_{i,j}\right)^2 },
\end{equation}
where $u_{i,j}$ is the numerical solution and $\bar{u}_{i,j}$ is the 
exact Sakovich' solution at time  $t=j\Delta t$.

\begin{figure}
\vspace*{2mm}
\psfig{file=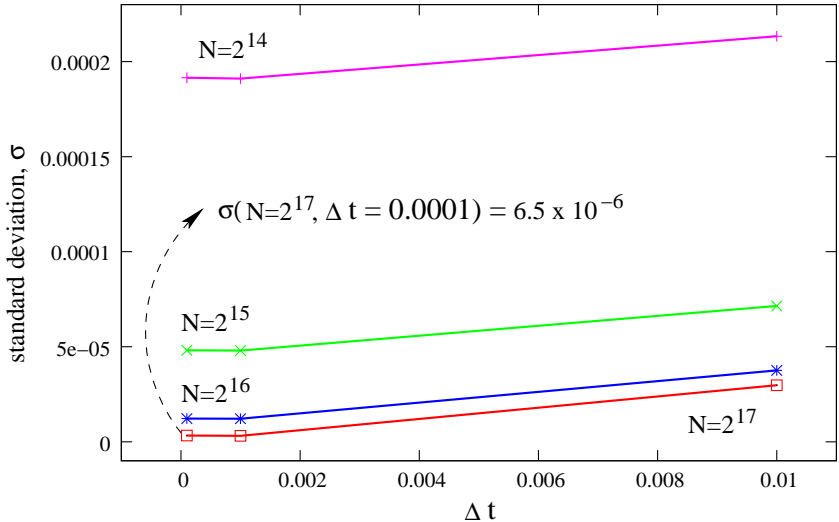,width=0.8\linewidth}
\caption{The dependence of the error of the polysymplectic integrator
from  $\Delta t$ for different values of $\Delta x$.}
\end{figure}

The results of the polysymplectic integration for different values of 
$\Delta t$
and $\Delta x = X_{max}/N$ ($X_{max} = 400$)  
are shown of Fig. 3. As expected, the error decreases with 
$\Delta x$ and $\Delta t$ decreasing.  
The polysymplectic method appears to be more effective 
than the pseudo-spectral method. 
For example, for $N=2^{17}$ and $\Delta t=0.0001$ the error 
of the polysymplectic scheme  $\sigma \approx 6.5 \times 10^{-6}$, 
while for the pseudo-spectral method $\sigma\approx 7 \times 10^{-5}$. 
The CPU time required by  the polysymplectic methods is 40000 sec, 
while the pseudo-spectral method requires $\approx 100000$ sec 
(on 3GHz Pentium 4 PC).

In conclusion, we have used the polysymplectic formulation of SPE 
in order to construct the geometric poly\-symplectic integrator of SPE. 
We have compared the effectiveness of the corresponding numerical scheme 
with the pseudo-spectral method which uses the  Runge-Kutta integration. 
The poly\-symplectic integration appears to be an order of magnitude 
more precise and approximately 2.5 times faster at long propagation times 
than  the pseudo-spectral method. 
A comparison with the exact solution of SPE shows that the  
poly\-symplectic integrator is more stable 
and robust than other schemes, and also preserves the energy functional.

\end{document}